\documentstyle[11pt,preprint,tighten,epsf,floats,rotating,aps]{revtex}
\preprint{CEBAF-TH-95-20}

\begin{document}
\thispagestyle{empty}
\title{HQET and Form Factor Effects in $B\to K^{(*)}\ell^+\ell^-$}
\author{W. Roberts}
\address{Department of Physics, Old Dominion University, Norfolk, VA
23529,
USA \\
and \\
Continuous Electron Beam Accelerator Facility \\
12000 Jefferson Avenue, Newport News, VA 23606, USA.}
\maketitle
\begin{abstract}
We examine the rates for the exclusive decays $B\to K^{(*)}\ell^+
\ell^-$. We use the scaling predictions of the
heavy quark effective theory to extract the necessary form factors
from fits to various combinations of data.
These data include the $D\to K^{(*)}\ell\nu$ semileptonic decays, as
well as the nonleptonic decays $B\to K^{(*)}\psi^{(\prime)}$
and the rare decay $B\to K^*\gamma$. We use different
parametrizations of form factors, and find that integrated decay
rates are not very sensitive to the forms chosen. However, the decay
spectra and the forward-backward asymmetry
in $B\to K^*\ell^+\ell^-$ are sensitive to
the forms chosen for the form factors, while the lepton polarization
asymmetry in
$\bar B^0\to \bar K^0\mu^+\mu^-$ is largely independent of the choice
of form factors.
Contributions from charmonium resonances dominate the spectra and
integrated rates.
In our `best' scenario, we find $Br(\bar B^0\to \bar K^0\mu^+\mu^-)=
2.0\pm 0.3\times 10^{-6}$ and $Br(\bar B^0\to \bar K^{*0}\mu^+\mu^-)=
8.1\pm 2.0\times 10^{-6}$. We also make predictions for other
polarization observables in these decays.
\end{abstract}
\pacs{  {\tt$\backslash$\string pacs\{13.25.Hw, 1.39.Fe, 14.40.Lb,
14.40.Nd \}} }
\newpage
\setcounter{page}{1}
\def\slash#1{#1 \hskip -0.5em / }
\setcounter{page}{2}

\section{Introduction}

The rare dileptonic and radiative decays of $B$ mesons have been the
subject of much recent interest. This is because the
operators responsible for these decays are absent in the standard
model at tree level, and first appear at one-loop level.
As a result, these decays can provide sensitive tests of many issues,
both within and beyond the standard model. The mass
of the top quark and the Higgs boson, the existence or not of other
Higgs multiplets, right-handed massive gauge bosons, or even extra
left-handed massive gauge bosons, as well as questions concerning
supersymmetric models are just some of the issues to which these
decays are sensitive
\cite{fls,ali1,ali,stn,rgra,rgrb,rgr1,rgr2,rgr,pyr,rdi,cglo,ahm,liu,mnl}.

In order for these issues to be probed with any kind of precision in
these decays, it is crucial that all of the long-distance effects
be understood. At present, it is believed that this is the case for
inclusive processes such as $B\to X_s\ell^+\ell^-$, the rates for
which are
taken to be the rates for the corresponding free-quark process. In
this regard, the operator-product-expansion (OPE) of the
heavy quark effective theory (HQET) has been used to treat
inclusive decays beyond the free-quark approximation
\cite{fls,ali1,chay,inclusive}. This approximation is actually the
leading term in a systematic expansion
in the inverse of the $b$-quark mass, and becomes arbitrarily
accurate as the mass of the $b$ quark approaches infinity. In
addition, it has been
shown that corrections to the free-quark picture first arise at order
$1/m_b^2$, so that the predictions for the inclusive decay rates are
expected to be quite reliable \cite{chay}.

There are, however, two regions of phase space in which the OPE of
HQET may be less reliable in predicting the inclusive decay rates
\cite{fls}. The first is near the charmonium
resonances, as the matrix elements of the four-quark operators that
contribute in this region may be subject to large final state
interactions. These may be beyond
the scope of the HQET treatment of the inclusive process. The second
is in the corner of phase space where $P_{X_s}^2\approx m_s^2$, where
$P_{X_s}$ is the four-momentum
of the hadronic final state $X_s$. This essentially arises from the
fact that, for the free quark decay, the spectral end-point occurs at
$P_{X_s}^2=m_s^2$, while for
the case of real hadrons, it occurs at $P_{X_s}^2=m_K^2$. Apart from
this, it is believed that the OPE of HQET provides a reliable
description of the inclusive decays.

For the exclusive decays, the situation is not quite as rosy, as the
free quark operators of the inclusive processes are replaced by
hadronic matrix elements, which are described in terms of a number of
{\it a priori} unknown, uncalculable, non-perturbative form factors.
The dependence of these form factors on the appropriate kinematic
variable may be modeled, but this muddles things as it introduces
some
model dependence in the extraction of information from the measured
quantities.

In this regard, one may use the predictions of the heavy quark
effective theory (HQET)
\cite{IW1,BG,HG1,FGGW,FG,MRR1,ML,FGL,MRR2,IW2,HG2,MN,iw2,hqet,wrfl}
to relate the form factors for the exclusive rare
decays of $B$ mesons to those of the semileptonic decays of $D$
mesons. There are two possible problems with this approach. The first
is
that the charm quark is not particularly heavy, and application of
HQET to the decays of charmed mesons may be of questionable validity
and value. The second is that to apply the form factors for the $D$
decays to $B$ decay processes requires extrapolation of the form
factors well
beyond the range that is kinematically accessible in $D$ decays.

Despite the relative `lightness' of the $c$ quark, the predictions of
HQET appear to be validated experimentally. For instance, the
predictions
for the decays of the $\Lambda_c$ \cite{MRR2,MRR3} are supported by
experimental measurements \cite{argus1,argus2}. In addition, and
perhaps more
importantly, the predictions of HQET for the decays $B\to D\ell\nu$,
in which the charm quark is treated as heavy, appear to be supported
by experimental
data. One may expect this success to carry over to the decays of
charmed mesons, thus justifying the use of HQET for such decays.

The question of extrapolation of form factors is a delicate one. In a
recent article, Roberts and Ledroit \cite{wrfl} have shown that
depending on the choice of form factor parametrizations, as well as
on the choice of form factor parameters, the form factors for $D$
decays may be applied with or without success to $B$ decays. The
question of success or non-success was a crucial one for the
nonleptonic decays $B\to K^{(*)}\psi^{(\prime)}$, for which the
question of factorization or not of the matrix element is also of key
importance.
Similar results have been reported by other authors
\cite{gourdin,leyaouanc,carlson}.

In \cite{wrfl}, the authors found that all of the data treated,
namely $D\to K^{(*)}\ell\nu$, $B\to K^{(*)}\psi^{(\prime)}$ and
$B\to K^*\gamma$, could be described in terms of a single set of
universal form factors. In this article, we use the results of that
work to analyse the decays $B\to K\ell^+\ell^-$ and $B\to K^*\ell^+
\ell^-$ in some detail, but concentrate on form factor effects rather
than the
effects of QCD coefficients, as these have been treated elsewhere by
many authors. In the case of the latter process, we also examine
the forward-backward asymmetry. In \cite{wrfl}, effects due to
charmonium resonances, and charm and light continua, were ignored.
These
are included in the present analysis.

The rest of this article is organized as follows. In the next section
we discuss the standard model effective Hamiltonian for the
rare dileptonic decays of interest, as well as the form factors for
the exclusive decays, and their HQET relations to the form factors
for the semileptonic decays of $D$ mesons. Our results for the total
decay rates, spectra, forward-backward asymmetries
and lepton polarization asymmetries are presented in
section III, and section IV presents our conclusions.

\section{Effective Hamiltonian And Form Factors}

\subsection{Rare Decays}

In the standard model, the effective Hamiltonian for the decay $b\to
s\ell^+\ell^-$ has the form
\begin{eqnarray}
{\cal H}_{\rm eff}&=&\frac{G_F}{\sqrt{2}}\frac{\alpha}{4
\pi}V_{ts}^*V_{tb}
\left[
2i\frac{m_b}{q^2} C_7(m_b)\bar{s}\sigma_{\mu\nu }q^\nu (1+\gamma_5)b
\bar \ell\gamma^\mu \ell\right.\nonumber\\
&+&\left.\vphantom{\frac{4 G_F}{\sqrt{2}}} C_9(m_b)\bar{s}\gamma_\mu
\left(1-\gamma_5\right)b\bar{\ell}\gamma^\mu\ell
+C_{10}(m_b)\bar{s}\gamma_\mu\left(1-\gamma_5\right)b\bar{\ell}
\gamma^\mu\gamma_5\ell\right]\label{effham},
\end{eqnarray}
where the Wilson coefficients $C_i(m_b)$ are as in the article by
Buras {\it et
al.} \cite{rgrb}. We choose not to reproduce these coefficients here:
the interested
reader may consult the rich literature on this subject. We do point
out, however, that $C_9$ and $C_{10}$
receive short distance contributions from the continua of light and
charm $q\bar q$ pairs, as well
as from charmonium resonances ($C_9$ only). This latter may be
thought of as arising from the nonleptonic decay
$B\to K^{(*)}\psi$, followed by the leptonic decay of the charmonium
vector resonance, $\psi\to\ell^+\ell^-$.
Thus, including these requires some assumption about the $B\to
K^{(*)}\psi$ amplitude.

As has been done by other authors, we assume that this amplitude can
be treated in the factorization approximation,
so that the contribution from each charmonium vector resonance $V$
can be written as
\begin{equation}
C_9^V=\frac{16\pi^2}{3}\frac{V_{cb}V^*_{cs}}{V_{tb}V^*_{ts}}\left(
\frac{f_V}{m_V}\right)^2\frac{a_2}{q^2-m_V^2+im_V\Gamma_V}.
\end{equation}
Here, $m_V$ is the mass of the charmonium state, $\Gamma_V$ is its
width, and $f_V$ is its decay constant. The constant
$a_2$ is the phenomenological factorization constant, whose absolute
value has been measured to be about 0.24. The
sign of $a_2$ is still uncertain, so we explore the effects of
changing this sign in the results that we present.

The hadronic matrix elements of the operators in eqn. (\ref{effham})
are
\begin{eqnarray}\label{ffgeneral}
\left<K(p^\prime)\left|\bar s\gamma_\mu c\right|B(p)
\right>&=&f^B_+(p+p^
\prime)_\mu+f^B_-(p-p^\prime)_\mu,\nonumber\\
\left<K(p^\prime)\left|\bar s\gamma_\mu \gamma_5c\right|B(p)
\right>&=&0,\nonumber\\
\left<K^*(p^\prime,\epsilon)\left|\bar s\gamma_\mu c\right|B(p)
\right>&=&ig^B\epsilon_{\mu\nu\alpha\beta}
\epsilon^{*\nu}(p+p^\prime)^\alpha(p-p^\prime)^\beta,\nonumber\\
\left<K^*(p^\prime,\epsilon)\left|\bar s\gamma_\mu\gamma_5 c
\right|B(p)\right>&=&f^B\epsilon^*_\mu+a^B_+\epsilon^*\cdot p(p+p^
\prime)_\mu
+a^B_-\epsilon^*\cdot p(p-p^\prime)_\mu,\nonumber \\
\left<K(p^\prime)\left|\bar s\sigma_{\mu\nu} b\right|B(p)
\right>&=&is^B
\left[\left(p+p^\prime\right)_\mu\left(p-p^\prime\right)_\nu-
\left(p+p^\prime\right)_\nu\left(p-p^\prime\right)_\mu\right],
\nonumber\\
\left<K^*(p^\prime,\epsilon)\left|\bar s\sigma_{\mu\nu} b\right|B(p)
\right>&=&\epsilon_{\mu\nu\alpha\beta}\left[g^B_+\epsilon^{*\alpha}
\left(p+p^\prime\right)^\beta
+g^B_-\epsilon^{*\alpha}\left(p-p^\prime\right)^\beta\right.\nonumber
\\
&&\left.+h^B\epsilon^*\cdot p\left(p+p^\prime\right)^\alpha\left(p-p^
\prime\right)^\beta\right].
\end{eqnarray}
Due to the relation
\begin{equation}
\sigma^{\mu\nu}\gamma_5=\frac{i}{2}\varepsilon^{\mu\nu\alpha\beta}
\sigma_{\alpha\beta},
\end{equation}
we can easily relate the matrix elements involving $\sigma_{\mu\nu}$
to those in which the
current is $\bar s\sigma_{\mu\nu}\gamma_5b$. The superscripts $B$ on
the form factors
signify that they are the ones appropriate to the decays of the $B$
mesons. These form
factors may be related to the corresponding ones for decays of $D$
mesons, using the
predictions of HQET.

The full formalism of HQET as it applies to these decays has been
presented in \cite{wrfl}. Here, we briefly present
the salient points of the discussion. In HQET, a heavy $B$ meson
traveling with velocity $v$ is represented by the
Dirac matrix \cite{falk}
\begin{equation}
B(v)\to \frac{1+\slash{v}}{2}\gamma_5.\nonumber
\end{equation}
The matrix elements of interest are then \cite{wrfl,MR}
\begin{eqnarray} \label{ff1}
\left<K(p)\left|\bar s\Gamma h_v^{(c)}\right|{\cal B}(v)\right>&=&{
\rm Tr}\left\{\left(\xi_1+\slash{p}\xi_2\right)\gamma_5\Gamma
\frac{1+\slash{v}}{2}\gamma_5\right\},\nonumber\\
\left<K^*(p,\epsilon)\left|\bar s\Gamma h_v^{(c)}\right|{\cal B}(v)
\right>&=&{\rm Tr}\left\{\left[\left(\xi_3+\slash{p}\xi_4\right)
\epsilon^*\cdot v+\slash{\epsilon}^*\left(\xi_5+\slash{p}\xi_6\right)
\right]\Gamma
\frac{1+\slash{v}}{2}\gamma_5\right\},
\end{eqnarray}
where
\begin{equation}
\left|B(v)\right>=\sqrt{m_B}\left|{\cal B}(v)\right>.
\end{equation}
These $\xi_i$ are independent of the masses of the heavy quarks and
mesons, as well as of the exact form
of the Dirac matrix $\Gamma$. Thus, they are valid for both $D\to
K^{(*)}$ and $B\to K^{(*)}$, as well as for
transitions mediated by vector, axial-vector and tensor currents.

The relationships between the form factors of eqn. (\ref{ffgeneral})
and the $\xi_i$ are
\begin{eqnarray}\label{xitof}
\xi_1&=&\frac{\sqrt{m_B}}{2}\left(f^B_++f^B_-\right),\nonumber\\
\xi_2&=&\frac{1}{2\sqrt{m_B}}\left(f^B_--f^B_+\right)=-\sqrt{m_B} s^B,
\nonumber\\
\xi_3&=&\frac{m_B^{3/2}}{2}\left(a^B_++a^B_-\right),\nonumber\\
\xi_4&=&\frac{\sqrt{m_B}}{2}\left(2g^B-a^B_++a^B_-
\right)=m_B^{3/2}h^B,
\nonumber\\
\xi_5&=&-\frac{1}{2\sqrt{m_B}}\left(f^B+2m_Bv\cdot p g^B\right)=-
\frac{
\sqrt{m_B}}{2}\left(g^B_++g^B_-\right),\nonumber\\
\xi_6&=&\sqrt{m_B}g^B=\frac{1}{2\sqrt{m_B}}\left(g^B_--g^B_+\right).
\end{eqnarray}
The corresponding relationships for $D$ meson
form factors require the replacement of all factors of $m_B$ in eqn.
(\ref{xitof}) by factors of $m_D$.
Finally, we note that inclusion of radiative corrections requires the
replacement \cite{qcd}
\begin{equation}
\xi_i^{b\to s}=\xi_i^{c\to s}\left[\frac{\alpha_s(m_b)}{
\alpha_s(m_c)}\right]^{-\frac{6}{25}}.
\end{equation}

\section{Results And Discussion}

All of the results we present are obtained by using the form factor
parametrizations of \cite{wrfl}. In that work,
two scenarios were explored for the form factors. In the first
scenario, $\xi_1$ and $\xi_4$ had the form
\begin{equation}
\xi_i=a_i\exp{\left[-b_i\left(v\cdot p-m_{K^{(*)}}\right)^2
\right]}=a_i\exp{\left[-\frac{b_i}{4m_D^2}\left(q^2_{{\rm max}}-q^2
\right)^2\right]}\label{exp2},
\end{equation}
$\xi_2$ and $\xi_5$ had the form
\begin{equation}
\xi_i=a_i\exp{\left[-b_i\left(v\cdot p-m_{K^{(*)}}\right)
\right]}=a_i\exp{\left[-\frac{b_i}{2m_D}\left(q^2_{{\rm max}}
-q^2\right)\right]}\label{exp1},
\end{equation}
while $\xi_5$ and $\xi_6$ had the form
\begin{equation}
\xi_i=a_i\exp{\left[-b_i\left(v\cdot p\right)^2\right]}\label{exp3}.
\end{equation}
In the second scenario, the $\xi_i$ were parametrized as
\begin{equation}
\xi_i=a_i\left(1+b_iv\cdot p\right)^{n_i},
\end{equation}
with $n_i=-2,\,-1,\,0,\,1$.

In each scenario, the $a_i$ and $b_i$ were free parameters that were
fixed by fitting to various combinations
of experimental measurements. Thus, for each scenario, four sets of $
\{a_i,b_i\}$ were generated. These corresponded to fits to
(I) the semileptonic decays $D\to K^{(*)}\ell\nu$; (II) the
semileptonic decays $D\to K^{(*)}\ell\nu$ and the nonleptonic
decays $B\to K^{(*)}\psi$; (III) the semileptonic decays $D\to
K^{(*)}\ell\nu$, the nonleptonic decays $B\to K^{(*)}\psi$ and the
nonleptonic decays $B\to K^{(*)}\psi^\prime$; (IV) the semileptonic
decays $D\to K^{(*)}\ell\nu$, the nonleptonic decays
$B\to K^{(*)}\psi$, the
nonleptonic decays $B\to K^{(*)}\psi^\prime$ and the rare decay $B\to
K^* \gamma$. A fuller discussion of these fits and parameter sets is
given in
\cite{wrfl}, but we emphasize that all of the results we present are
obtained using form factors that are consistent at least with the
measurements
in $D\to K^{(*)}\ell\nu$, including polarization ratios. In addition,
in this analysis, we have used $V_{tb}=0.9988$, $V_{ts}=0.03$,
$V_{cs}=0.9738$,
$V_{cb}=0.041$, $m_b=4.9$ GeV, $m_c$=1.5 GeV, $m_t$=177 GeV.

\begin{figure}
\centerline{\mbox{\begin{turn}{0}%
\epsfxsize=3.0in\epsffile{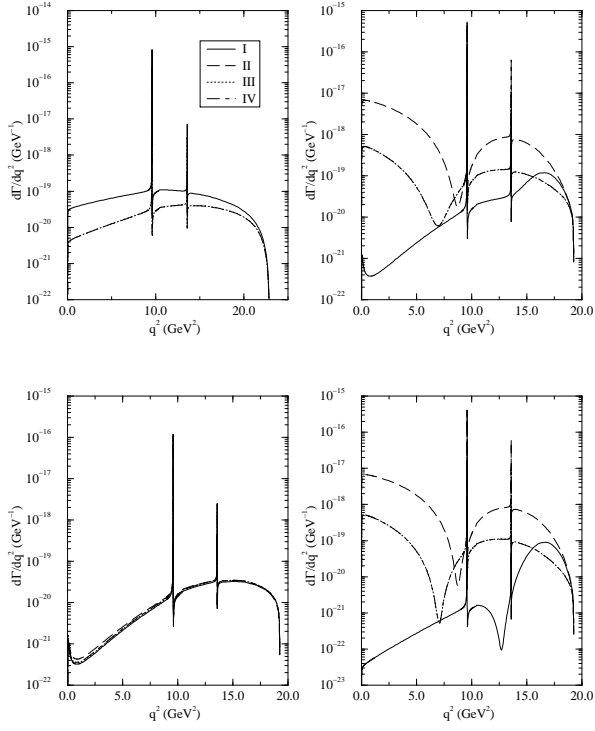}\end{turn}}}
\vskip 0.25in
\caption{Differential decay rates for the processes $B\to K\mu^+
\mu^-$ and $B\to K^*\mu^+\mu^-$, in the exponential
scenario. The graphs are, starting at the top left and moving
clockwise:
$B\to K\mu^+\mu^-$; $B\to K^*\mu^+\mu^-$; $B\to K^*\mu^+\mu^-$ for
longitudinally polarized $K^*$'s; $B\to K^*\mu^+\mu^-$ for
transversely polarized $K^*$'s. In each graph,
I means that the form factors used were obtained from a fit in which
only data for
 $D\to K^{(*)}\ell\nu$ have been included;
II means that data for $D\to K^{(*)}\ell\nu$ and $B\to K^{(*)}J/\psi$
have
been included; III means that data for $D\to K^{(*)}\ell\nu$, $B\to
K^{(*)}J/\psi$ and
$B\to K^{(*)}\psi^\prime$ have been included; IV
means that data for $D\to K^*\ell\nu$, $B\to K^*J/\psi$, $B\to K^*
\psi^\prime$ and
$B\to K^*\gamma$ have all been included, and does not apply to
the process $B\to K\mu^+\mu^-$.
\label{rarerates1}}
\end{figure}

\begin{figure}
\centerline{\mbox{\begin{turn}{0}%
\epsfxsize=3.0in\epsffile{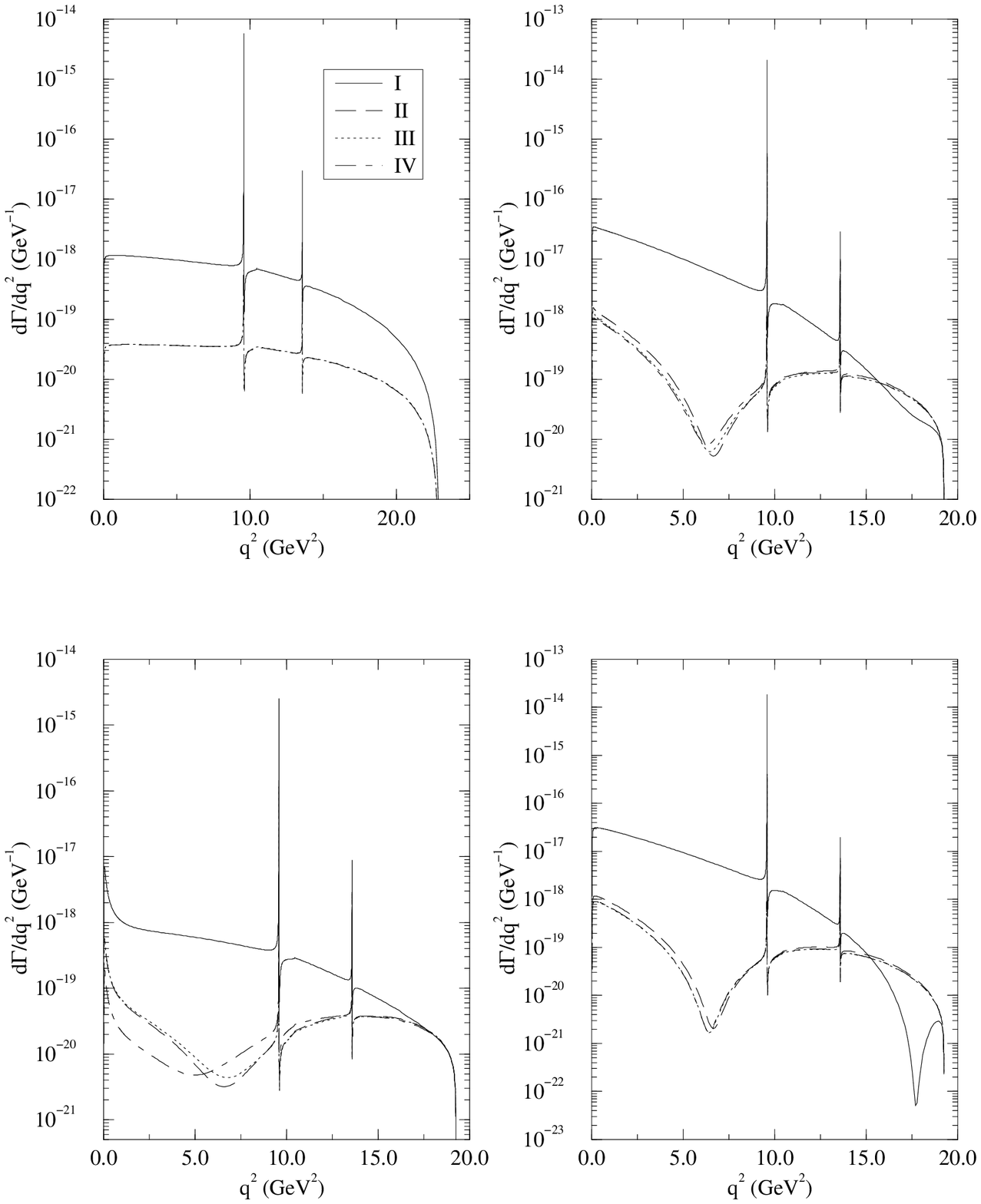}\end{turn}}}
\vskip 0.25in
\caption{Differential decay rates for the processes $B\to K\mu^+
\mu^-$ and $B\to K^*\mu^+\mu^-$, in the multipolar
scenario. The graphs are, starting at the top left and moving
clockwise:
$B\to K\mu^+\mu^-$; $B\to K^*\mu^+\mu^-$; $B\to K^*\mu^+\mu^-$ for
longitudinally polarized $K^*$'s; $B\to K^*\mu^+\mu^-$ for
transversely polarized $K^*$'s.
In each graph, the key is as in fig. \protect\ref{rarerates1}.
\label{rarerates2}}
\end{figure}

\begin{figure}
\centerline{\mbox{\begin{turn}{0}%
\epsfxsize=3.0in\epsffile{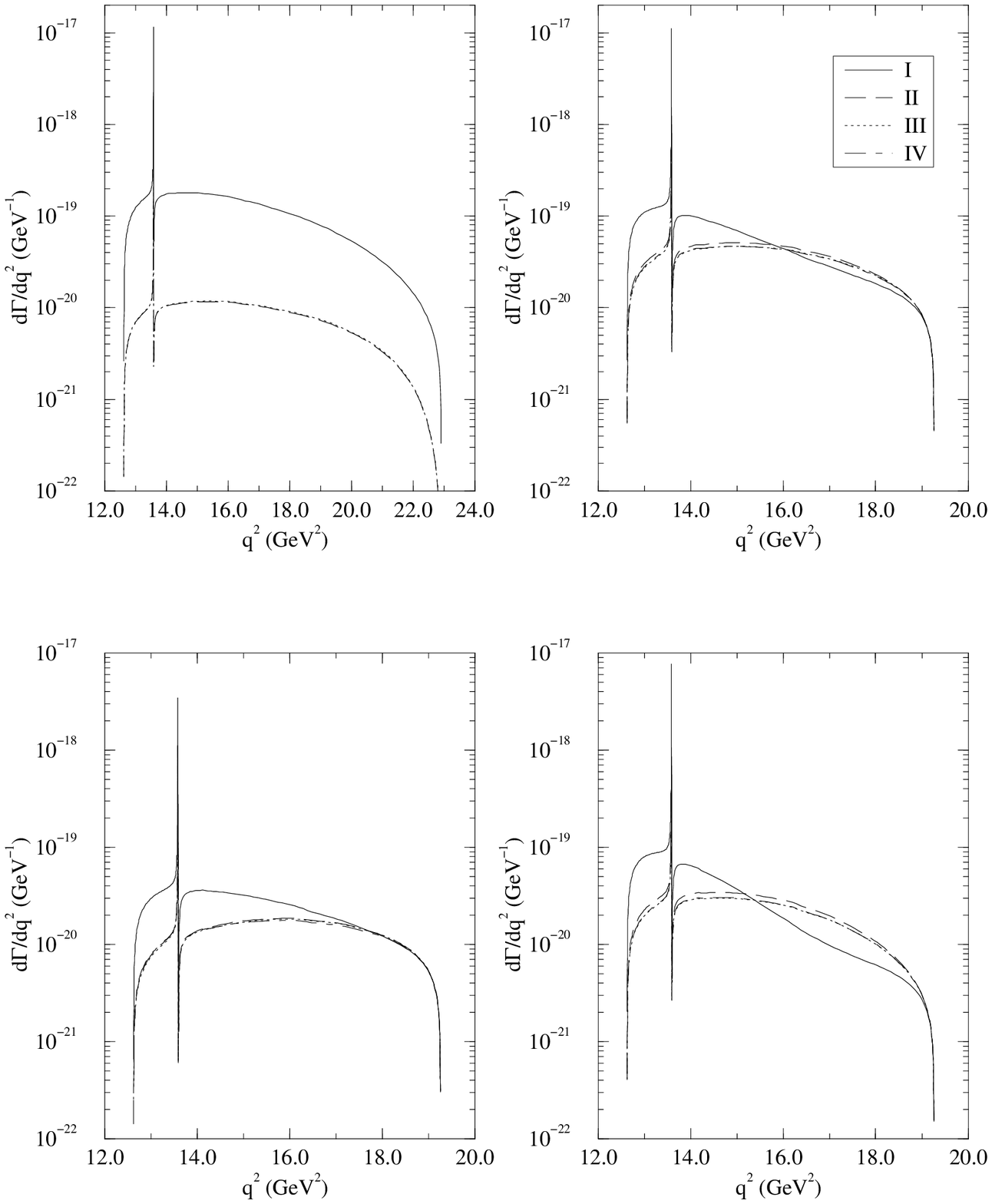}\end{turn}}}
\vskip 0.25in
\caption{Differential decay rates for the processes $B\to K\tau^+
\tau^-$ and $B\to K^*\tau^+\tau^-$, in the multipolar
scenario. The graphs are, starting at the top left and moving
clockwise:
$B\to K\tau^+\tau^-$; $B\to K^*\tau^+\tau^-$; $B\to K^*\tau^+\tau^-$
for
longitudinally polarized $K^*$'s; $B\to K^*\tau^+\tau^-$ for
transversely polarized $K^*$'s.
In each graph, the key is as in fig. \protect\ref{rarerates1}.
\label{rarerates3}}
\end{figure}

In fig. \ref{rarerates1} we show our results for the rare dileptonic
decays $B\to K^{(*)}\mu^+\mu^-$ using the form factors of the
exponential scenario.
Fig. \ref{rarerates2} shows the corresponding spectra obtained using
the form factors of the multipolar scenario. In each of figs.
\ref{rarerates1} and \ref{rarerates2},
the graph at the top left is $d\Gamma/dq^2$ for $B\to K\mu^+\mu^-$,
while the second upper graph
shows the spectrum for $B\to K^*\mu^+\mu^-$. The lower graphs show
the corresponding curves for transversely and longitudinally
polarized $K^*$'s
in $B\to K^*\mu^+\mu^-$. For comparison, Fig. \ref{rarerates3} shows
the corresponding spectra for production of $\tau$ leptons, in the
multipolar scenario.

 The most dominant features of these
curves are the sharp maxima due to the first two vector charmonium
resonances. Apart from these two features, the spectra we have
obtained are very
similar to those obtained in \cite{wrfl}. In particular, the zeroes
in some of the distributions still persist.

The two charmonium resonances also dominate the total rates, as the
numbers
in tables \ref{rare1} and \ref{rare2} are all at least twice as large
as the corresponding numbers reported in \cite{wrfl}, where the
resonance effects were not included. In these tables, the labels of
the columns correspond to the fits described above. This means, for
instance, that
the predictions of set III are
obtained using the form factors from fit III, in which we have
included the data for the semileptonic decays $D\to K^{(*)}\ell\nu$,
the nonleptonic decays
$B\to K^{(*)}\psi$ and the nonleptonic decays $B\to K^{(*)}\psi^
\prime$. The errors that we quote in all of the numbers we report are
estimates only,
and are obtained by using the covariance matrix that arises from the
fit.

\squeezetable

\begin{table}
\caption{Predictions for decay rates of $B\to K^{(*)}\mu^+\mu^-$ in
the exponential
scenario. I means that only $D\to K^{(*)}\ell\nu$ has been included
in the fit;
II means $D\to K^{(*)}\ell\nu$ and $B\to K^{(*)}J/\psi$ have been
included; III means $D\to K^{(*)}\ell\nu$, $B\to K^{(*)}J/\psi$ and
$B\to K^{(*)}\psi^\prime$ have been included; IV
means $D\to K^*\ell\nu$, $B\to K^*J/\psi$, $B\to K^*\psi^\prime$ and
$B\to K^*\gamma$ have all been included, and applies only to decays
with
$K^*$'s in the final state.
\label{rare1}}
\begin{tabular}{|l||c|c|c|c|c||}\hline
Quantity & Experiment & I & II & III & IV \\ \hline\hline
$\Gamma_{B\to K\mu^+\mu^-}$ ($10^{-18}$ GeV) & $<158.0$ & $2.32\pm
4.46$ & $0.78\pm 0.19$ & $0.78\pm 0.19$ & - \\ \hline
$\Gamma_{B\to K^*\mu^+\mu^-}^T$ ($10^{-18}$ GeV) & - & $0.39\pm 0.22$
& $0.42\pm 0.10$ & $0.41\pm 0.08$ & $0.41\pm 0.09$ \\ \hline
$\Gamma_{B\to K^*\mu^+\mu^-}^L$ ($10^{-18}$ GeV)  & - & $0.28\pm
0.07$ & $28.9\pm 20.3$ & $2.47\pm 0.32$ & $2.46\pm 2.65$ \\ \hline
$\Gamma_{B\to K^*\mu^+\mu^-}$ ($10^{-18}$ GeV) & $<10.1$ & $0.67\pm
0.22$ & $29.4\pm 20.3$ & $2.88\pm 0.28$ & $2.88\pm 2.65$ \\ \hline
\end{tabular}
\end{table}

\begin{table}
\caption{Predictions for decay rates of $B\to K^{(*)}\mu^+\mu^-$ in
the multipolar
scenario. The columns are as in table \protect\ref{rare1}.
\label{rare2}}
\begin{tabular}{|l||c|c|c|c|c||}\hline
Quantity & Experiment & I & II & III & IV \\ \hline\hline
$\Gamma_{B\to K\mu^+\mu^-}$ ($10^{-18}$ GeV) & $<158.0$ & $1.86\pm
1.59$ & $0.86\pm 0.15$ & $0.87\pm 0.15$ & - \\ \hline
$\Gamma_{B\to K^*\mu^+\mu^-}^T$ ($10^{-18}$ GeV) & - & $10.8\pm
2.18$ & $0.69\pm 0.14$ & $0.69\pm 0.17$ & $0.60\pm 0.06$ \\ \hline
$\Gamma_{B\to K^*\mu^+\mu^-}^L$ ($10^{-18}$ GeV)  & - & $142.8\pm
28.5$ & $3.54\pm 2.48$ & $2.86\pm 2.07$ & $2.93\pm 0.89$ \\ \hline
$\Gamma_{B\to K^*\mu^+\mu^-}$ ($10^{-18}$ GeV) & $<10.1$ & $153.6\pm
28.4$ & $4.22\pm 2.48$ & $3.56\pm 2.07$ & $3.52\pm 0.89$ \\ \hline
\end{tabular}
\end{table}

Apart from the charmonium features shown in these figures, the
differences in the predicted spectra for different parametrizations
of form factors,
but within the same scenario, and from the exponential to the
multipolar scenario, are quite striking. The reader is reminded that
for all of these
curves, the form factors are consistent with all of the measurements
in the semileptonic decays $D\to K^*\ell\nu$. Nevertheless, apart
from a few
obvious exceptions, the predictions for the total rates are
surprisingly similar for the different parametrizations and
scenarios.

\begin{figure}
\centerline{\mbox{\begin{turn}{0}%
\epsfxsize=3.0in\epsffile{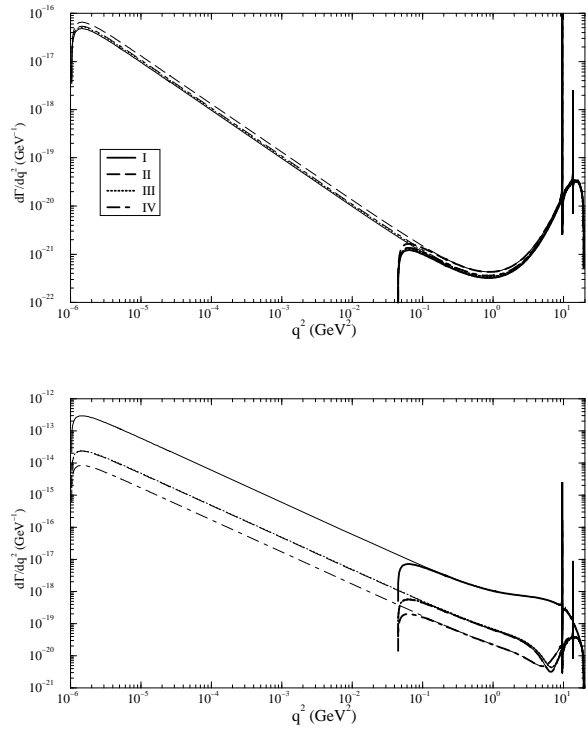}\end{turn}}}
\vskip 0.25in
\caption{Differential decay rates for the processes $B\to K^*\mu^+
\mu^-$ and $B\to K^*e^+e^-$, for transversely polarized $K^*$'s.
The exponential scenario is the upper graph, while the multipolar
scenario
is the lower graph. In each graph, the thick curves are for $B\to K^*
\mu^+
\mu^-$, while the thin curves are for $B\to K^*e^+e^-$, and the key
is as in fig. \protect\ref{rarerates1}.
\label{rarerates4}}
\end{figure}

If the final leptons are electrons, all of the curves we have shown
are essentially the same, with the exception of those for
transversely
polarized $K^*$'s for small $q^2$ (and consequently, for unpolarized
$K^*$'s as well). This is because the differential decay rate for
transversely
polarized $K^*$'s behaves like $1/q^2$ for small $q^2$, and the
different end-points for electrons and muons means that the spectra
are different at
small $q^2$. In fact, the $1/q^2$ dependence is softened by a factor
of $\sqrt{q^2-4m_\ell^2}$ in the decay rate. That phase space extends
further for
electron pairs has essentially no impact on the rate for $B\to K
\ell^+\ell^-$, nor for longitudinally polarized $K^*$'s in $B\to K^*
\ell^+\ell^-$. However,
there is a significant increase in the rate for transversely
polarized $K^*$'s, with a slightly less significant effect for
unpolarized $K^*$'s. This is seen
by comparing the numbers in tables \ref{rare2} and \ref{rare3}. The
effect is also shown in fig. \ref{rarerates4}. For tau leptons, all
rates are smaller by
about an order of magnitude.

\begin{table}
\caption{Predictions for decay rates of $B\to K^{(*)}e^+e^-$ in the
multipolar
scenario. The columns are as in table \protect\ref{rare1}.
\label{rare3}}
\begin{tabular}{|l||c|c|c|c|c||}\hline
Quantity & Experiment & I & II & III & IV \\ \hline\hline
$\Gamma_{B\to Ke^+e^-}$ ($10^{-18}$ GeV) & $<158.0$ & $1.87\pm
1.60$ & $0.86\pm 0.15$ & $0.87\pm 0.15$ & - \\ \hline
$\Gamma_{B\to K^*e^+e^-}^T$ ($10^{-18}$ GeV) & - & $15.7\pm
3.29$ & $1.07\pm 0.36$ & $1.09\pm 0.42$ & $0.74\pm 0.08$ \\ \hline
$\Gamma_{B\to K^*e^+e^-}^L$ ($10^{-18}$ GeV)  & - & $144.7\pm
28.9$ & $3.59\pm 2.52$ & $2.90\pm 2.11$ & $2.96\pm 0.91$ \\ \hline
$\Gamma_{B\to K^*e^+e^-}$ ($10^{-18}$ GeV) & $<10.1$ & $160.4\pm
28.7$ & $4.66\pm 2.53$ & $4.00\pm 2.14$ & $3.70\pm 0.90$ \\ \hline
\end{tabular}
\end{table}

In addition to the differential decay rate, there are two other
quantities of interest for these decays. One is  the differential
forward-backward
asymmetry, $A_{{\rm FB}}$, which may be defined as
\begin{equation}
A_{{\rm FB}}=\frac{\int^1_0\frac{d\Gamma}{dq^2d\cos{\theta_\ell}}d
\cos{\theta_\ell}-\int^0_{-1}\frac{d\Gamma}{dq^2d\cos{\theta_\ell}}d
\cos{\theta_\ell}}
{\int^1_0\frac{d\Gamma}{dq^2d\cos{\theta_\ell}}d\cos{\theta_\ell}+
\int^0_{-1}\frac{d\Gamma}{dq^2d\cos{\theta_\ell}}d\cos{\theta_\ell}}.
\end{equation}
Here, $\theta_\ell$ is the angle that the negatively charged lepton
makes, in the dilepton rest frame, with the momentum of the daughter
$K^*$, and the
denominator is simply $d\Gamma/dq^2$. This quantity is identically
zero, in the standard model, for $B\to K\ell^+\ell^-$.

\begin{figure}
\centerline{\mbox{\begin{turn}{0}%
\epsfxsize=3.0in\epsffile{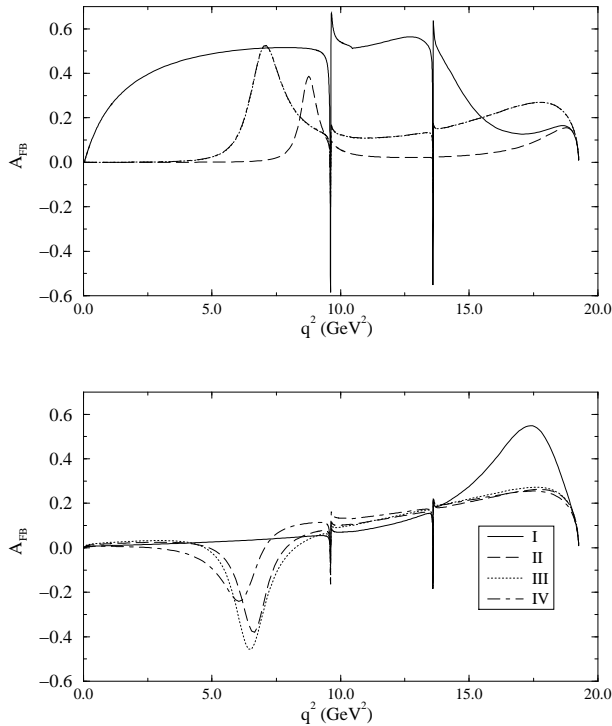}\end{turn}}}
\vskip 0.25in
\caption{The forward-backward asymmetry, $A_{{\rm FB}}$, in $B\to K^*
\mu^+\mu^-$. The upper graph is for the
exponential scenario, while the lower is for the multipolar
scenario. In each graph, the key is as in fig. \protect
\ref{rarerates1}.
\label{rarerates5}}
\end{figure}

\begin{figure}
\centerline{\mbox{\begin{turn}{0}%
\epsfxsize=3.0in\epsffile{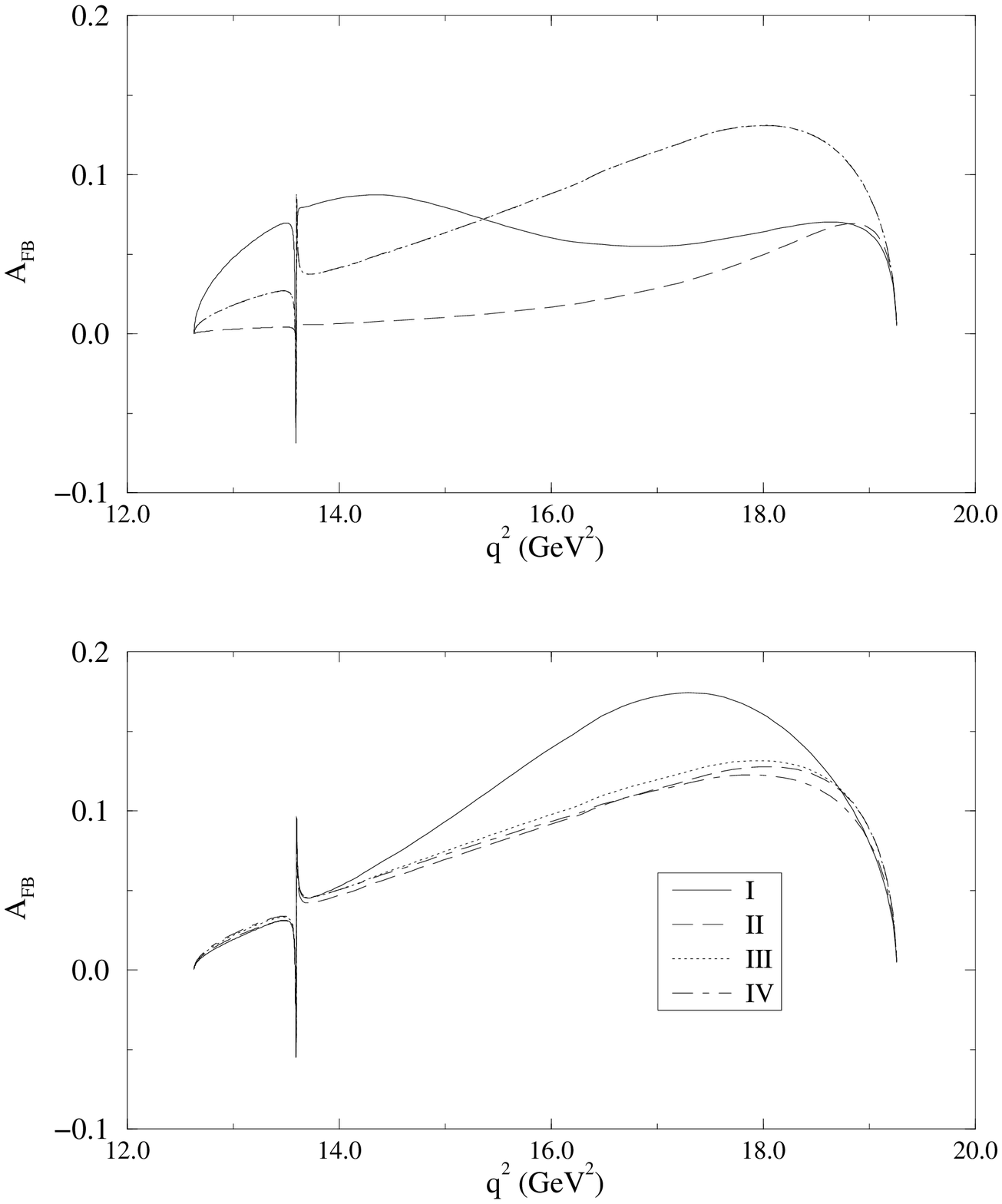}\end{turn}}}
\vskip 0.25in
\caption{The forward-backward asymmetry, $A_{{\rm FB}}$, in $B\to K^*
\tau^+\tau^-$. The upper graph is for the
exponential scenario, while the lower is for the multipolar
scenario. In each graph, the key is as in fig. \protect
\ref{rarerates1}.
\label{rarerates6}}
\end{figure}

The forward-backward asymmetries that result from our calculations
are shown in fig. \ref{rarerates5} for $B\to K^*\mu^+\mu^-$, and in
figure
\ref{rarerates6} for $B\to K^*\tau^+\tau^-$. In each case, the upper
graph is for the
exponential scenario,
while the lower one is for the multipolar one. We again emphasize
that the
differences in the curves for each graph arise from changes in the
parameters of the form factors. We also point out that the form of
this asymmetry will
also depend on the physics content of the Wilson coefficients, and
that the curves shown all correspond to standard-model physics only.

\begin{figure}
\centerline{\mbox{\begin{turn}{0}%
\epsfxsize=3.0in\epsffile{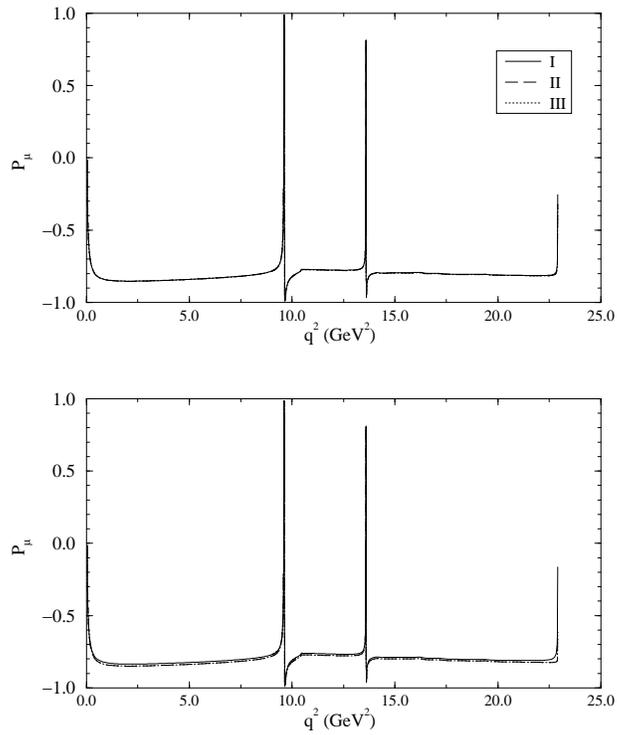}\end{turn}}}
\vskip 0.25in
\caption{The lepton polarization asymmetry, ${\cal P}_\mu$, in $B\to
K\mu^+\mu^-$. The upper graph is for the
exponential scenario, while the lower is for the multipolar
scenario. In each graph, the key is as in fig. \protect
\ref{rarerates1}. The curves for the different fits are essentially
indistinguishable on this scale.
\label{rarerates7}}
\end{figure}

\begin{figure}
\centerline{\mbox{\begin{turn}{0}%
\epsfxsize=3.0in\epsffile{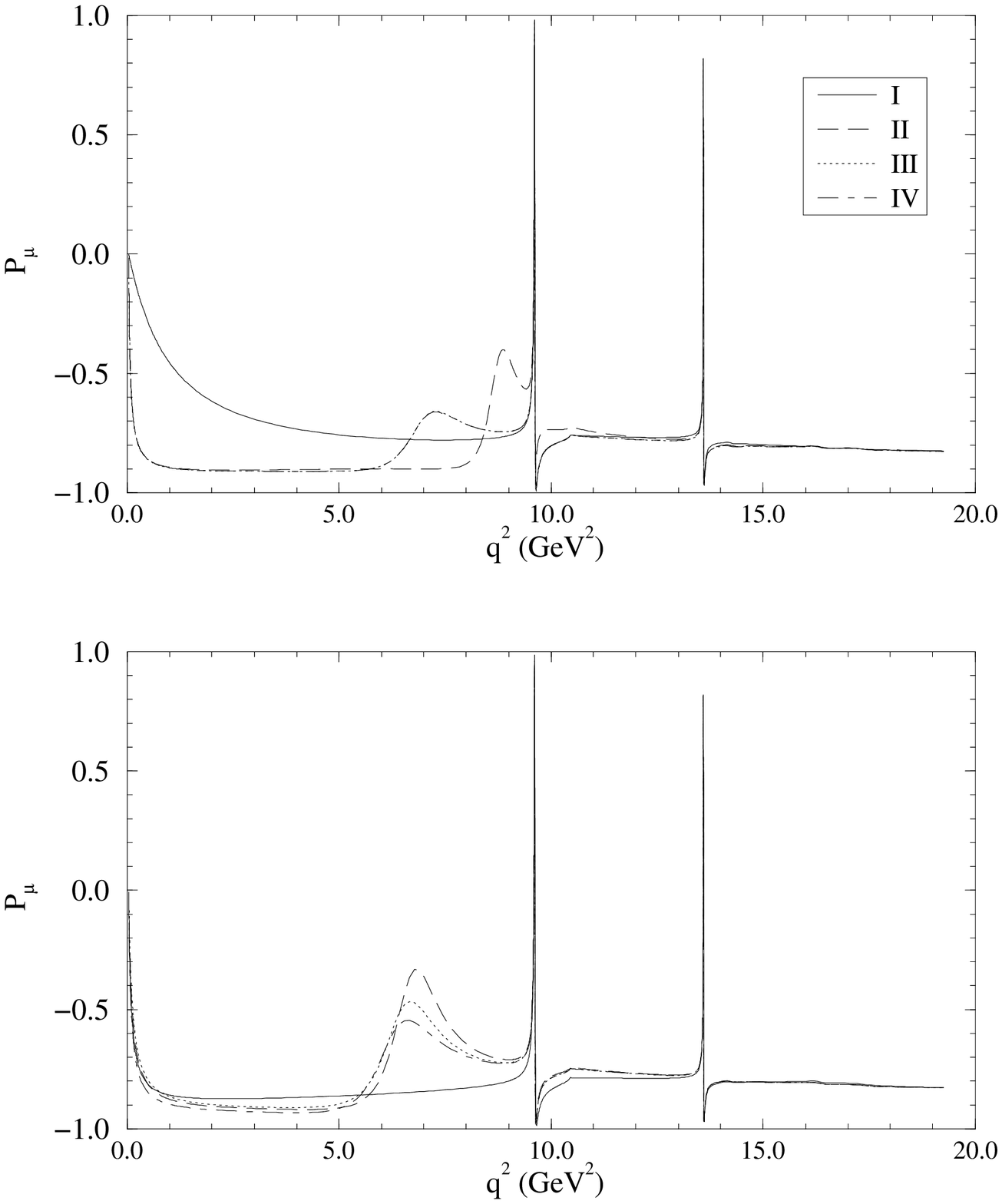}\end{turn}}}
\vskip 0.25in
\caption{The lepton polarization asymmetry, ${\cal P}_\mu$, in $B\to
K^*\mu^+\mu^-$. The upper graph is for the
exponential scenario, while the lower is for the multipolar
scenario. In each graph, the key is as in fig. \protect
\ref{rarerates1}.
\label{rarerates8}}
\end{figure}

\begin{figure}
\centerline{\mbox{\begin{turn}{0}%
\epsfxsize=3.0in\epsffile{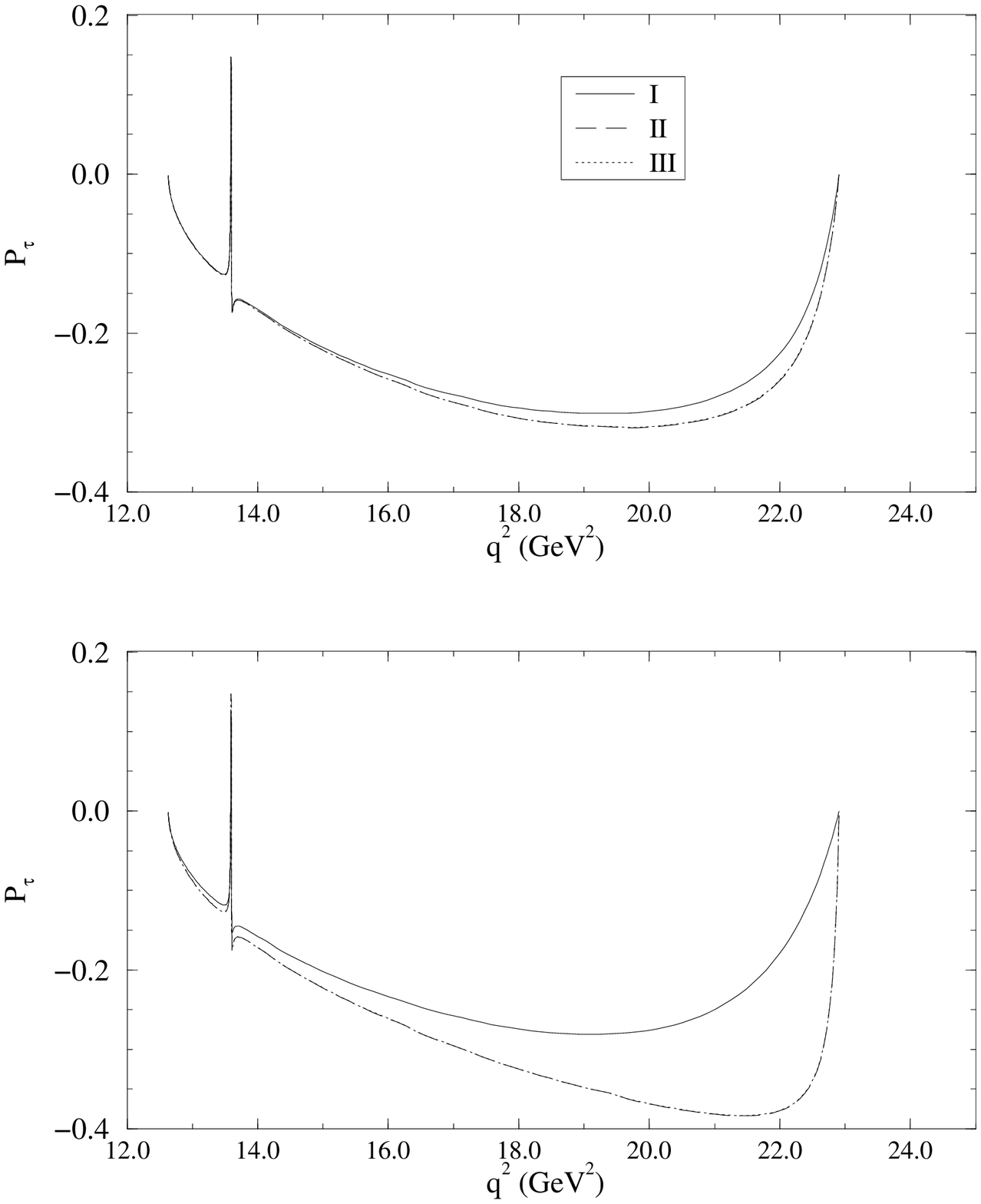}\end{turn}}}
\vskip 0.25in
\caption{The lepton polarization asymmetry, ${\cal P}_\tau$, in $B\to
K\tau^+\tau^-$. The upper graph is for the
exponential scenario, while the lower is for the multipolar
scenario. In each graph, the key is as in fig. \protect
\ref{rarerates1}.
\label{rarerates9}}
\end{figure}

\begin{figure}
\centerline{\mbox{\begin{turn}{0}%
\epsfxsize=3.0in\epsffile{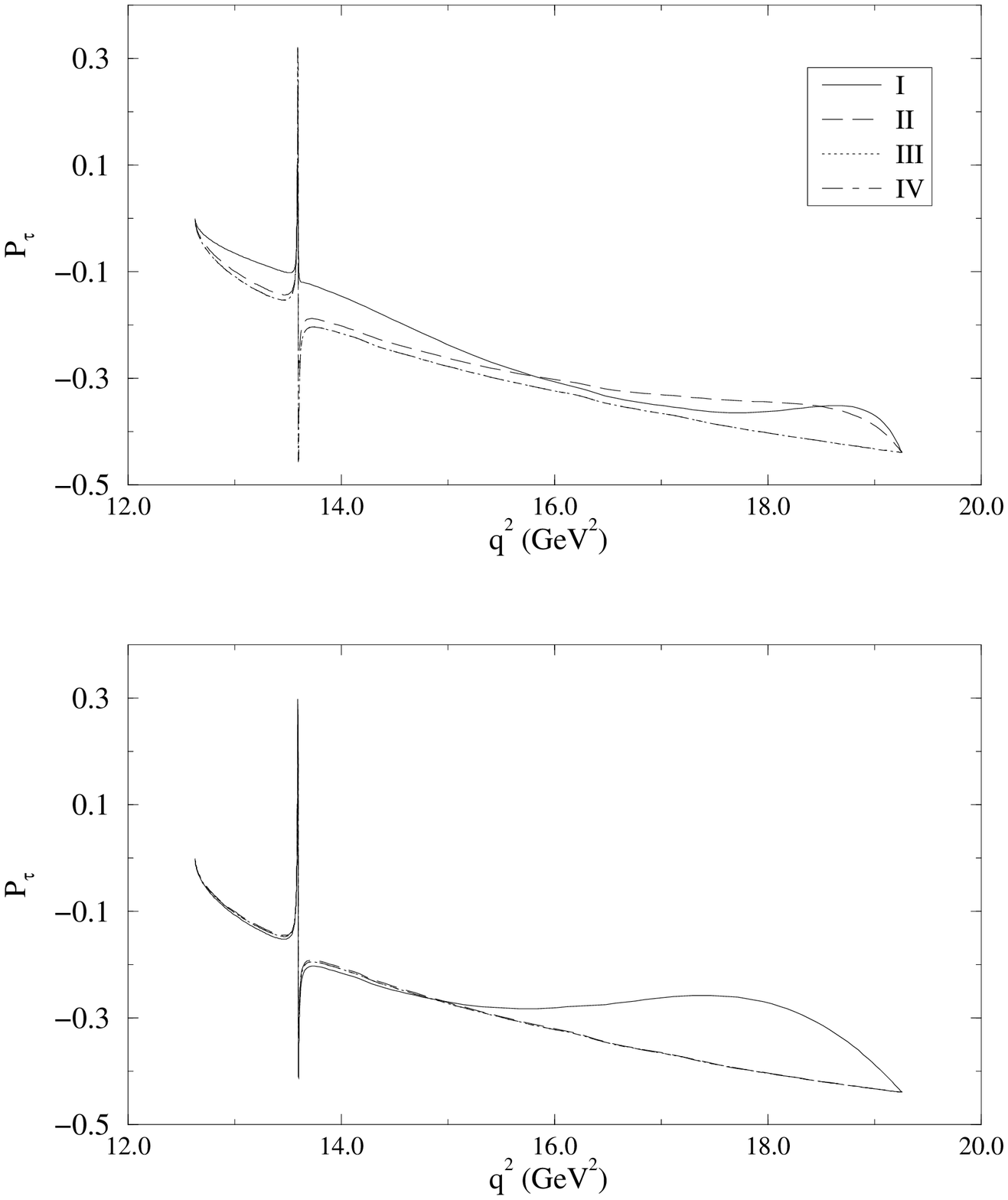}\end{turn}}}
\vskip 0.25in
\caption{The lepton polarization asymmetry, ${\cal P}_\tau$, in $B\to
K^*\tau^+\tau^-$. The upper graph is for the
exponential scenario, while the lower is for the multipolar
scenario. In each graph, the key is as in fig. \protect
\ref{rarerates1}.
\label{rarerates10}}
\end{figure}

The second quantity of interest in these decays is the lepton
polarization asymmetry, defined as
\begin{equation}
{\cal P}_\ell=\frac{\left.\frac{d\Gamma}{dq^2}\right|_{\lambda=-1}-
\left.\frac{d\Gamma}{dq^2}\right|_{\lambda=+1}}
{\left.\frac{d\Gamma}{dq^2}\right|_{\lambda=-1}+\left.\frac{d
\Gamma}{dq^2}\right|_{\lambda=+1}},
\end{equation}
where the subscripts $\lambda$ denote whether the spin of the $
\ell^-$ is alligned parallel ($\lambda=+1$) or antiparallel ($
\lambda=-1$)
to its motion. Fig. \ref{rarerates7} shows the results we obtain for
this quantity for muons in $B\to K\mu^+\mu^-$, while fig.
\ref{rarerates8}
shows the corresponding results for $B\to K^*\mu^+\mu^-$. Figs.
\ref{rarerates9} and \ref{rarerates10}, respectively, show the
corresponding
results for tau leptons.

The most striking feature of fig. \ref{rarerates7} is the
insensitivity of ${\cal P}_\mu$ to the parametrization of the form
factors. The same feature
also appears in fig. \ref{rarerates8}, but mainly for the large
dilepton mass region of phase space. The insensitivity of this
polarization
observable to form factors has not previously been anticipated as far
as we know, and suggests that
the polarization asymmetry could be one of the more useful
observables for examining the physics content of the Wilson
coefficients.

This asymmetry in $B\to K\ell^+\ell^-$ is independent of form factor
parametrizations due to a combination of two effects.
The first of these is the small lepton mass (for $\ell=\mu$ or $e$),
which means that many terms in the differential decay rate are small
for most regions of phase space. The second is the
relative smallness of
the $C_7$ coefficient compared with $C_9$ and $C_{10}$. The
consequence of this, together with the small lepton mass, is that any
form factor dependence in the polarization asymmetry disappears. In
fact, to a very good approximation, in the limit in which $C_7$ is
small, we find
\begin{equation}\label{polarization}
{\cal P}_\mu\approx 2\frac{{\rm Re}C_9C_{10}^*}{\left|C_9\right|^2+
\left|C_{10}\right|^2}+{\cal O}\left(C_7\right).
\end{equation}
This is also independent of the assumptions of HQET, since only the
hadronic vector and axial vector operators contribute to
${\cal P}_\mu$: eqn. (\ref{polarization}) does not rely on any
special relationships among form factors.
This asymmetry therefore provides a direct measure of the
interference between $C_9$ and $C_{10}$. In addition, experimental
observation
of significant departures from this nearly constant value for muons
would signal larger values of $C_7$, and therefore, possibly, new
physics.

Figure \ref{rarerates8} shows a similar effect in the polarization of
the muons produced in $B\to K^*\mu^+\mu^-$, particularly at large
values of the
dilepton mass. In fact, to the same level of approximation, the
lepton polarization in this process is given by the same expression,
eqn. (\ref{polarization}).
This is a better approximation at large values of $q^2$, as form
factor effects become more significant at smaller $q^2$ for this
decay.

Unfortunately, in the case of $\tau$ leptons, where the polarization
may be more easily measured, the fact that the lepton mass is large
means
that this polarization variable depends on the particular choice of
form factors, as can be seen in figs. \ref{rarerates9} and
\ref{rarerates10}. Nevertheless, some
simplification does occur at the kinematic end-point, where
$q^2=q^2_{\rm max}$. There, form factor dependence again
disappears, and the tau polarization asymmetry is determined solely
in terms of the coefficients $C_9$ and $C_{10}$ (assuming that $C_7$
is small),
and the hadron and lepton masses, $m_B$, $m_{K^*}$ and $m_\tau$ (at
this kinematic point in $B\to K\ell^+\ell^-$, the polarization
asymmetry vanishes identically).
Thus, for given values of the Wilson coefficients, there is a firm
prediction for
this asymmetry at maximum $q^2$ in $B\to K^*\tau^+\tau^-$.
We note that
the form of the curve we obtain for this quantity in the exclusive
channel $B\to K^*\tau^+\tau^-$ is very similar to that obtained by
Hewett \cite{hewett}
in the inclusive process $B\to X_s\tau^+\tau^-$.

\begin{figure}
\centerline{\mbox{\begin{turn}{0}%
\epsfxsize=3.0in\epsffile{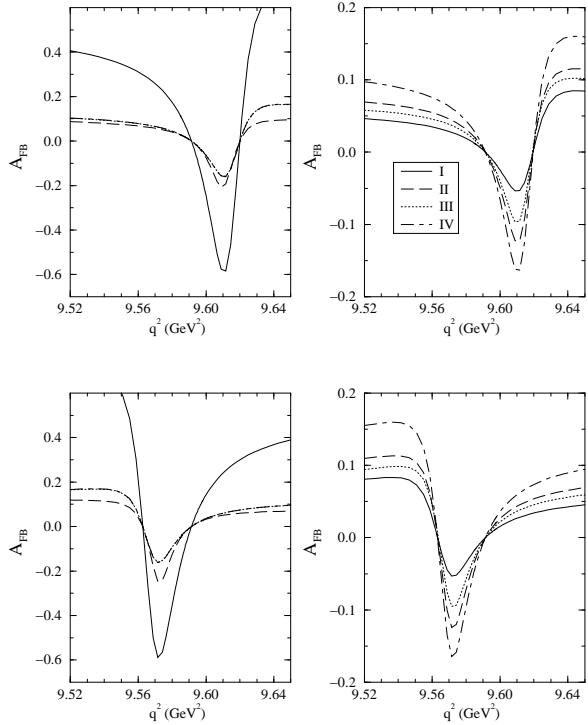}\end{turn}}}
\vskip 0.25in
\caption{The effect of the sign of $a_2$ on $A_{{\rm FB}}$ in $B\to
K^*\ell^+\ell^-$. The two graphs on the left are
for the exponential scenario, while the two on the right are for the
multipolar scenario. The two upper graphs are for $a_2>0$, while the
two
lower ones are for $a_2<0$. In each graph, the key is as in fig.
\protect\ref{rarerates1}.
\label{rarerates11}}
\end{figure}

Finally, we turn to the question of the sign of $a_2$. Since this
parameter enters only through the charmonium resonances, it should
not be surprising that
the effects of a change in its sign are most clearly visible in the
vicinity of these resonances. In the decay spectra, there is some
modification of the
shape, but only very close to each resonance. The effect on $A_{{\rm
FB}}$ is a little more interesting, and
is displayed in fig. \ref{rarerates11}. However,
since the difference between the two sets of curves shown occurs
between $q^2$ of 9.52 and 9.64 GeV$^2$, it is doubtful whether future
experiments
will ever have the $q^2$ resolution needed to distinguish one set of
curves from the other. Thus, we would suggest that the prospects of
determining the
sign of $a_2$ from these decays are not very promising. We find a
similar result when we examine the effect of the sign of $a_2$ on the
lepton
polarization asymmetry.

Our predictions for the process $B\to K\ell^+
\ell^-$ are two to three orders of magnitude smaller than present
experimental upper limits, but they are about three times as large as
the rates predicted by
Ali {\it et al.} \cite{ali}. Our absolute rates correspond to
branching fractions of $(1.8\pm 0.4)\times 10^{-6}$ in the
exponential scenario, and $(2.0\pm 0.3)\times 10^{-6}$ in the
multipolar scenario.

For $B\to K^*\ell^+\ell^-$ our predicted branching fractions are
$(6.6\pm 0.8)\times 10^{-6}$ and $(8.1\pm 2.0)\times 10^{-6}$
in the exponential and multipolar scenarios, respectively, for muon
pairs. For electron pairs, the multipolar scenario predicts a
branching fraction of $(8.5\pm 2.1)\times 10^{-6}$. Furthermore, we
find the ratio $\Gamma_T/\Gamma_L$ in $B\to K^*\mu^+\mu^-$
to be $0.17\pm 0.06$ in the exponential scenario and $0.20\pm 0.08$
in the multipolar scenario. For $B\to K^*e^+e^-$, the
multipolar scenario predicts a value of $0.25\pm 0.10$ for this
quantity. It is somewhat surprising but nonetheless reassuring that
even this
polarization ratio is largely independent of form factor
parametrizations. This suggests that our predictions for total rates
should be quite reliable, as
uncertainties due to form factor parametrizations have less impact on
integrated quantities.

The numbers that we have quoted for $B\to K^*\ell^+\ell^-$
correspond to III of the exponential scenario and IV of the
multipolar scenario. In the case of the exponential scenario, we have
chosen
III as the best numbers to present for two related reasons. The first
is that the theoretical uncertainties on IV are unreasonably large,
while those on III are more `reasonable'. However, as can be seen
from the graphs (and the tables), there is very little difference
between III and IV
in this scenario. The problem arises because in going from III to IV,
we have added the CLEO measurement of $B\to K^*\gamma$ to the fit,
and the exponential scenario can not accomodate the experimental
measurement (the `best fit' in this scenario
is more than a factor of 100 smaller than the measurement).
Consequently, the fit parameters (and our predictions) remain
the same in going from III to IV, but the errors on the predictions
have increased.
In contrast with this, IV of the multipolar scenario provides a
satisfactory description of all the data used in the fit, including
the
measured rate for $B\to K^*\gamma$.

\section{Conclusion}

There is a plethora of issues that we have not touched in this note.
Extensions to the standard model and their effects
on the Wilson coefficients, scale dependence of these coefficients,
and the forms of these coefficients at leading order and beyond
are beyond the scope of this article. While these issues are very
important, recent calculations suggest that, at least for the
inclusive
decays, some kind of convergence is at hand. This is not so for the
exclusive decays. Our results indicate that while results for
integrated
rates and lepton polarization asymmetries appear to be largely
independent of the parametrization chosen for the form factors,
differential rates and the forward-backward
asymmetry are not. Measurements of these quantities in exclusive
channels will therefore serve to probe form factor models or
parametrizations.
This is therefore similar to the situation in the exclusive decay $B
\to K^*\gamma$, which has turned out to be a testing ground for
form factor models.

The scenario that best describes all of the experimental data is the
multipolar one and, in this scenario, we find that the universal
form factor $\xi_6$ is linear in $v\cdot p$. Using this scenario,
we predict $Br(\bar B^0\to \bar K^0\mu^+\mu^-)=
(2.0\pm 0.3)\times 10^{-6}$ and $Br(\bar B^0\to \bar K^{*0}\mu^+
\mu^-)=
(8.1\pm 2.0)\times 10^{-6}$. These numbers are consistent with other
model calculations \cite{pyr}, and include the effects of the first
two charmonium vector
resonances. We also predict $\Gamma_T/\Gamma_L$ in
$\bar B^0\to \bar K^{*0}\mu^+\mu^-$ to be $0.20\pm 0.08$.

In the course of this study we have discovered that the polarization
asymmetries in the decays $B\to K^{(*)}\mu^+\mu^-$ are, to a very
good
approximation, independent of form factor effects, and are determined
solely in terms of the Wilson coefficients $C_9$ and $C_{10}$.
This is particularly so for the decays to the ground state kaons, as
the approximation is valid over all of phase space. Thus, these
observables could be very useful tools for probing the physics
content of the Wilson coefficients. However, in order for this to be
a practical tool, experimentalists must be able to measure the
polarization of the daughter muons in these decays, with adequate
precision.

Hewett \cite{hewett} suggests that the polarization of the tau
leptons could be measurable at $B$ factories that are under
construction. If that
is the case, there should certainly be sufficient numbers of events
produced in the muon channels, at least in the `clean' region away
from
the two charmonium resonances, as the decay rates for muons and taus
are comparable in this region of phase space. The remaining question
is therefore
simply one of whether the polarization of the muon can be measured in
these decays. This may be possible for sufficiently slow muons, or if
the
muons can be stopped in the detector.

For tau leptons, simplifications such as those mentioned above do not
occur, and the polarization asymmetry depends on form factors for
almost
all of phase space. The sole exception is at the kinematic end point
in the decay $B\to K^*\tau^+\tau^-$, when the dilepton pair has
maximum $q^2$.
There, for given values of the Wilson coefficients, there is a firm
prediction for
this asymmetry in $B\to K^*\tau^+\tau^-$. We emphasize again that
the fact that the asymmetry is independent of form factors does not
depend on the assumptions of the heavy quark effective theory.
Whether either
of these polarization effects can ever be measured will have to await
completion of the $B$ factories.

\acknowledgements

Special thanks are due F. Ledroit for continued interest and
encouragement.
The author gratefully acknowledge helpful conversations with D.
Atwood, N. Isgur and C. Carlson, as well as
the support of the National Science
Foundation under grant PHY 9457892, and the U.
S. Department of Energy under contracts DE-AC05-84ER40150 and DE
FG05-94ER40832.


\end{document}